# The Function of Gesture in an Architectural Design Meeting

Willemien Visser
LTCI (Laboratoire commun en Traitement et Communication de l'Information), UMR 5141 CNRS-TELECOM ParisTech
INRIA (National Institute for Research in Computer Science and Control, France)
willemien.visser@TELECOM-ParisTech.fr

**Abstract**
This text presents a cognitive-psychology analysis of spontaneous, co-speech gestures in a face-to-face architectural design meeting (A1 in DTRS7). The long-term objective is to formulate specifications for remote collaborative-design systems, especially for supporting the use of different semiotic modalities (multi-modal interaction). According to their function for design, interaction, and collaboration, we distinguish different gesture families: representational (entity designating or specifying), organisational (management of discourse, interaction, or functional design actions), focalising, discourse and interaction modulating, and disambiguating gestures. Discussion and conclusion concern the following points. It is impossible to attribute fixed functions to particular gesture forms. "Designating" gestures may also have a design function. The gestures identified in A1 possess a certain generic character. The gestures identified are neither systematically irreplaceable, nor optional accessories to speech or drawing. We discuss the possibilities for gesture in computer-supported collaborative software systems. The paper closes on our contribution to gesture studies and cognitive design research.

## 1. Introduction: analysis from a cognitive-psychology viewpoint as first stage in a cognitive-ergonomics analysis

This text presents a cognitive-psychology analysis of a design meeting, focusing on the function of gestures in collaborative design. We are interested by all the gestures that designers may use to elaborate on their project. This choice excludes, from our analysis, only signed gestures, commonly used instead of speech by deaf or hard of hearing people to convey meaning.

To examine the function of the gestures, we analysed them obviously in their articulation with other semiotic modalities (Goodwin 2002). However, we do discuss neither gesture-related expression forms such as posture, facial expression, gaze, or paralinguistic and prosodic aspects of speech, nor hand-made actions such as drawing. This is for reasons of space. Given the scarce knowledge on gesture in collaborative design, the exclusive description of gesture already provides a quantity of material that requires a sharp selectivity.

Our investigation is the first stage of a larger cognitive-ergonomics analysis. Its long-term objective is to identify the consequences that knowledge on gestures in face-to-face design meetings has for remote collaborative-design environments, especially with respect to multi-modal interaction modalities and their support.

The present analysis continues the approach adopted throughout our previous work in cognitive design research since the 1980s (Visser 2006a). We do not adopt, as a reference for analysis, pre-existing theoretical entities such as the design "task" or "process" invoked in design methods and prescriptive models of design. Instead, we focus on the dynamic aspects of the design activity as designers implement it during their actual work on a design project. Glock (this volume) adopts a comparable focus, analysing the way in which design is done in practice.

## 2. Gesture in collaborative designing

After a short introduction to gesture research (Kendon 2004, McNeill 2000), we present previous work on gesture in design meetings.

### 2.1 Gesture research

Gesture studies have been conducted, especially, in the fields of semiotics, ethology, dance and choreographic research, psychology / psycholinguistics, and pragmatics. They have focused mainly on everyday conversation; the use of gesture in goal-oriented, professional activities has hardly been studied, but there are some notable exceptions, such as studies by Heath and other members of the Work, Interaction and Technology research group (e.g., Heath and Luff 2007, Hindmarsh and Heath 2000), Goodwin (e.g., Goodwin 2003, Koschmann et al. 2006), and Mondada (2002). In cognitive ergonomics of design, analysis of gesture is at its beginnings.

### 2.2 Gesture in design meetings: previous studies

Compared with verbal and graphical interaction, gestural expression has barely been analysed in collaborative design.

In an early study of collaborative design (a remote control handset), Tang (1991) identified that the "hand-made actions" of writing, freehand drawing, and gesturing had three functions: "store information", "express ideas," and "mediate interaction." A full third of hand-made actions were gestures. Comparatively, these served hardly any information-storage function, while expression of new ideas was performed as much by gesture as by drawing and writing. However, more than half of all gestures served interaction mediation, considerably more than did other hand-made actions.

Bekker, Olson, and Olson (1995) distinguished four types of gesture in the design of an automatic post office: kinetic (action execution—which we do not qualify as gesture), spatial (indication of distance, location, or size), pointing, and other (emphasis on verbal utterances and attracting the attention). As purposes for gesture use, they identified design, management, and overall conversation regulation. All combinations of types of gestures and functions occurred.

Studying a team of architects, Murphy (2005) analysed their designing in terms of "collaborative imagining." He noted the use of gesture for imagining characteristics of design entities (their motion, structure, and functioning) and of user experience (actions performed on the design entities).

In our previous gesture-related research, we have developed a description language for graphico-gestural design activities (Détienne and Visser 2006, Détienne, Visser and Tabary 2006, Visser and Détienne 2005). We have used this language, combined with COMET, our method for analysing collaborative design (Darses et al. 2001), to describe and analyse the verbal and graphico-gestural dimensions of the interaction in an architectural design meeting. We have interpreted co-designers' graphico-gestural actions according to their functional roles in the project or the meeting (Détienne et al. 2006). We have also examined different forms of multimodal articulation between graphico-gestural and verbal modalities in parallel interactions between the designers: activities taking place at the same time revealed alignment or disalignment between the designers regarding the focus of their activities (Détienne and Visser 2006, Visser and Détienne, 2005).

We wish to emphasise two functions of gesture shown by these previous studies. (1) Gesture offers specific possibilities to render spatial (especially 3D) and motion-related qualities of entities, and to embody action sequences through their mimicked simulation (cf. Tversky and Lozano 2006). (2) Gesture plays an important "interactional" or, as we qualify it below, "organisational" role.

## 3. Data analysis: use of gesture by designers in an architectural meeting

In order to continue and broaden our previous work on gesture in collaborative architectural design, we chose the architectural meetings among the DTRS7 material. Given the explorative nature of our study and the time-consuming character of gesture analysis, we selected only one meeting, the first Architectural Meeting (A1).

### 3.1 Data analysed: design in the Architectural Meeting A1

The three participants in the meeting were Adam, the architect in charge of the project, and two clients, Anna, registrar of the cemetery, and Charles (whose precise function was not explicit); a DTRS7 organiser observed A1. The project concerned the design of a crematorium (or "chapel"), to be built on a site with already another crematorium.

Even if different roles can be distinguished among the participants, we consider that all three participants are doing design. We analyse design not as the activity of somebody whose profession is "designer," but as a cognitive activity. "Design consists in specifying an artifact (the artifact product), *given requirements* that indicate—generally neither explicitly, nor completely—one or more functions to be fulfilled, and needs and goals to be satisfied by the artifact, under certain conditions (expressed by constraints). At a cognitive level, this specification activity consists of constructing (generating, transforming, and evaluating) representations of the artifact *until they are so precise, concrete, and detailed* that the resulting representations—the 'specifications'—specify explicitly and completely the implementation of the artifact product" (Visser 2006a, p. 116).

### 3.2 Description scheme: distinguishing gestures according to their function in the design meeting

Not every hand or arm movement is a gesture. There is a difference between gestures and actions (Grosjean and Kerbrat-Orecchioni 2002). Actions have a practical aim: they transform the material state of the world and are, in principle, without any meaning, strictly speaking (even if bodily actions can also be analysed as contributing to the construction of meaningful structures). Gestures, on the contrary, convey meaning: they have a symbolic function and transform the cognitive state of the addressee.

The gestures analysed here are (1) "gesticulations" (Kendon 2004), that is, spontaneous, speech-accompanying gestures (McNeill 2000), and (2) "emblems", quasi-linguistic, lexicalised gestures, with conventional forms and meanings, and which are not necessarily speech-accompanying.

There are many different classifications of gestures. The most well-known ones, adopted by many authors, are those proposed by Kendon (2004) and by McNeill (2000). These classifications combine form and use of gestures. Two classical examples are deictics—pointing, surrounding, or covering gestures, generally considered to designate entities—and iconics—gestures that bear a physical resemblance with an entity, generally considered to convey the entity through this relationship. Our analysis is guided by the functions of the activities that the gestures are performing or supporting, not by their form. As we will show, a particular type of gesture may take various forms, and a particular gestural movement may fulfil different functions. A gesture may be "iconic" in McNeill (2000)'s classification, but we ask ourselves: *why* does a designer use this gesture *in this meeting*? Table 1 presents the five gesture families and subfamilies distinguished in this text as a result of our analysis (detailed in the next section).

Table 1. Families of gestures distinguished in this text (*see text)

| *main families of gestures* | *subfamilies* | *sub-subfamilies* |
|---|---|---|
| representation | | |
| | designation | |
| | | identification |
| | | qualification |
| | | comparison |
| | specification | |
| organisation | | |
| | discourse and interaction management | |
| | | management of one's own discourse |
| | | management of co-participants' interaction |
| | functional design-action management | |
| focalisation | | |
| modulation | | |
| disambiguation | | |

Since we analyse design as the construction of representations (Visser 2006a), our representational gestures correspond to Bekker, Olson, and Olson (1995)'s "design," and Murphy (2005)'s "imagining" activities. Naively, one might expect that, in order to develop the representation of the artefact, designers elaborate on its qualities. They do so indeed, but our observation has shown that they also often explicitly introduce the design entity concerned by their representational activities (object, process, action, state, or one of its components). We distinguish such designational activities from the representational activities that (further) specify the entity.

Designating gestures point out an entity (the *designatum)* to one's co-participants. Their form is most frequently a deictic. The designation may be more or less global: deictics vary from pointing with a finger to waving with a hand. Designata can be various: static physical objects, directions or movements in space, or moments in time. Generally considered the most typical gestures, deictics have received much attention, both in pragmatics and in computer-system and HCI research (but it is also the main type of gesture discerned by Glock, this volume, one of the rare DTRS7 colleagues referring occasionally to gesture).

In order to gesturally specify an entity, people bodily display one or more of its qualities. They may use illustrative gesticulations (Cosnier and Vaysse 1997) or emblems. Contrary to the deictic designating gesture, an illustrative gesture resembles its designatum: one must look closely at the gesture, because the designatum is—at least, partially—"in" it.

As most interaction between co-participants in a design meeting is or serves design in a more or less direct manner (Visser 2006a), we do not set apart interactional activities. Yet, we distinguish their management. Different authors have analysed and qualified management gestures in different ways: qualifications adopted are, for example, "interaction mediation" (Tang 1991), or "management" and "overall conversation regulation" (Bekker et al. 1995). We use the term "organisational" to refer to gestures that cover, on the one hand, the management of one's own discourse and of the interaction between the co-participants, and, on the other hand, the planning and organisation of the functional design actions (e.g., to propose to proceed first to brainstorming and only afterwards to the formulation of critique).

To organise their own discourse, people often use "beats." Beats are gesticulations whose form is shaped by one or two hands that move along with the rhythm of speech (as a music director conducting her orchestra). They structure discourse through their accentuating certain parts of it. A particularity that distinguishes beats from both pointing designating and illustrative gestures is that they tend to have the same form regardless of the discourse content.

Three types of gesture are not specific to collaborative design, as they can be found in any interaction: focalisation, modulation, and disambiguation. These functions can be combined with the previous ones.

Focalisation serves to make one's addressees focus on a fraction of the world that one considers critical—be it "outside" or in one's discourse. It always combines with one or more other functions. Each representational gesture focuses. Below, it will be presented as it combines with designation and with organisation.

Modulation is an expressive function: discourse or interaction components can be modulated, that is, emphasised, generally with an emotional loading.

Gestures are not simple illustrations of verbal discourse. They are generally associated with speech, but have their own contribution. They can also help to disambiguate a discourse or interaction component expressed using another semiotic system.

Our description scheme is preliminary with respect to "the" inventory of gestures used in A1, the analysed meeting—if ever such an exhaustive inventory might exist. We did not describe all gestures, but rather tried to identify the different types of gestures performed during A1, according to their function for design, interaction, and collaboration. Therefore, our analysis does provide us neither with an exhaustive list, nor with quantitative results.

## 4. Results: gestures made by the design co-participants

After an explanation of the formalism used in the data-extract examples, we present the results of our analysis.

### 4.1 Formalism used in the examples

We use the verbal protocol provided by the DTRS7 organisers, which provided a transcription cut into numbered lines. We will describe the physical realisation of certain gestures ("gesture movements"), introducing two or three lines under the transcription line. The first of these lines indicates, between square brackets, the identity of the gesturer and the duration of the gesture by reference to that of the corresponding portion of the verbal protocol, for example [g_Anna......]. The second and, if necessary, third line(s) present(s) a brief description of the gesture movement (see Extract 1).

Extract 1, A1, Formalism used for describing gestures

| | | |
|---|---|---|
| 352 | Anna | … they can just go in side by side but it's difficult to squeeze in to |
| 353 | | put the coffins on at the moment even because you've also you've got the |
| | | [g_Anna..............][g_Anna............ |
| | | *draws apart her hands* |
| | | *opens each hand* |
| 354 | | two catafalques in side by side and you need to have four routes for |
| | | ........................................][g_Anna.......................................... |
| | | *into a ball-enclosing shape* *spreads apart two fingers of each hand* |
| 355 | | people to go either you need the one in the middle for both people to go |
| | | ....................] |
| | | *and advances them over two separate parallel tracks* |

In Extract 1, saying "even because you've," Anna draws apart her hands. Then, during "also you've got the two catafalques in side by side", she opens each hand in a grasping movement, as if she encloses a ball with each hand; finally, saying "and you need to have four routes for people to go," she spreads apart two fingers of each hand and advances with these fingers over two separate parallel tracks.

### 4.2 Representational gestures: designation and specification

**Designation: an identificatory, qualificatory, or comparative orientation, combined with a focusing goal**. We observed that designating gestures generally are identificatory: gesturers designate entities in order for their co-participants to be able to identify them. Gestures were used in a qualificatory manner—even if qualification generally relied more on intonation, facial expression, or posture than on gesture. They could also be used comparatively: in contrast to "that chapel" (first pointing gesture), "the original concept of this chapel" (second pointing gesture) "was that the bier… would actually come out and meet the hearse" (A1, 245-246). Such comparative use was generally combined with an identificatory function. All three functions combined with a focusing use of the gestures in question.

**Spatial deictics.** In order to appreciate our impression, when first viewing A1, that spatial deictic gestures were quite frequent, we made a quick analysis of the first hour of the meeting. Out of the 215 verbally expressed spatial deictics, more than 200 indeed seemed to be accompanied or completed by a deictic gesture.

We observed the use of spatial deictics for various indications: indirectly *designating an entity* through *pointing to the location* (point or area) where the entity was situated, or directly *designating a direction* or *an itinerary*.

In Extract 2, Adam indicates three possible routes for entering the cremator area, tracing with a finger over the plan three lines following three different paths.

Extract 2, A1, Spatial deictic gesture designating three itineraries

| | | |
|---|---|---|
| 1134 | Adam | obviously there are numerous ways of getting into this accommodation |
| 1135 | | that's route number one the second route is round the end of the pond |
| | | [g_Adam.......................................................................... |
| | | *tracing with a finger consecutively three lines over the plan* |
| 1136 | | and the third route is through the chapel so there's numerous ways of |
| | | ......................................................] |

**Specify an entity: display its qualities.** Some examples of qualities used to gesturally specify entities were size, movement, luminousness, boldness, importance, intimate character, private character, calmness, or spiritual atmosphere.

Indeed, both physical and abstract qualities were specified, the first through "iconic", the second using "metaphoric" illustrative gestures—even if this distinction is a relative one, gestures rather being more or less iconic (Krauss, Chen and Gottesman 2000, p. 276). Furthermore, it is not the form of a gesture, but its relationship with the designatum that makes the gesture iconic or metaphoric. Iconic gestures bear a physical resemblance, metaphoric gestures a metaphoric relation with the entity they are meant to convey.

Adam's lining gestures in Extract 2 were iconic and metaphoric, both the routes and the gazing orientation being represented as lines.

In Extract 3, Adam moves apart his arms to ask Anna for the width necessary to make two coffins pass through a passage in the chapel.

Extract 3, A1, Gesture specifying a physical quality

| | | |
|---|---|---|
| 368 | Adam | OK so my question for you is how wide would it need to be for two |
| 369 | | coffins or if we're going for two it would need to be |

In Extract 4, Adam proposes a new door and possibly a new window for the AV technicians to be able to monitor the funeral services that precede and follow the one in progress. He points to the location where he plans this door and draws it. Charles, following with his finger the direction of view, gesturally "sees" what becomes possible.

Extract 4, A1, Gesturally "seeing" a possible direction of view

| | | |
|---|---|---|
| 635 | Adam | the answer is then to have a door there |
| 636 | Charles | a door |
| 637 | Adam | maybe a window |
| 638 | Charles | a window |
| 639 | Adam | and they can |
| 640 | Charles | and a window this way |

**Asserting a quality or asking for one**. Instead of gesturally attributing a quality to an entity, a participant may also gesturally enquire about it. In Extract 3, Adam asks his client for the required width of a passage.

In A1, request has been identified in combination with specification. Obviously, other representational and organisational expressions also can be combined with these two enunciative modalities (see below, e.g., Charles questioning his co-participants concerning seating).

### 4.3 Organisational gestures

In her answer to Adam who asks if indeed she does not see any need for alteration (A1, 299), Anna organises—and underlines (see Modulating gestures below)—her discourse with beats (Extract 5). To do so, she performs a sequence of hands opening and closing movements, accompanied with a hands rising and/or advancing movement. Considering as one beat a sequence of hands opening and closing, combined with a rising and/or advancing movement, Anna performs some 13 beats during this 5 seconds utterance.

Extract 5, A1, Organising one's discourse using beats

| | | |
|---|---|---|
| 300 | Anna | not no as long as there's a that's what they're I mean if we have a sort of |
| 301 | | consultation meeting with them that's what they will be interested in |

Interaction-management gestures may moderate co-participants' turn-taking by a stop-sign (an emblem) or call co-participants' attention by waving a hand, or lifting a finger. An example is Charles who, in order to focus the attention of his co-participants on a question concerning seating, points to the seats (A1, 392: "Are we going to have fixed seating?").

The next example may be used to illustrate several qualities that—not necessarily all coupled, as is the case here—may be encountered in gestural (or verbal) expression.

After a long intervention by Anna (A1, 174-194), Adam refocuses topics (A1, 197: "yes + so having got this this far we've got a bigger waiting room loos"). Pointing with his pen to the waiting room, that is, the design entity concerned in his coming discourse, he establishes a transition from Anna's discourse to the next topic that he presents: "having got this far."

- Adam uses two associated deictic expressions (gestural and verbal) to "point" "where [they] are reached in the meeting," that is, to make a designation in the notional domain.
- Adam's gesture precedes the associated verbal deictic expression "this far:" indeed, gestures do not necessarily follow their associated verbal expressions.
- There are other occurrences of organisationally used deictic expressions in the meeting (A1, 90-91: "this leads us to the first query I have"; A1, 323: "OK so having got this far").
- Adam's gesture is also an emphatic modulating gesture—something useful for an expression with an organisational purpose.
- Besides its interaction-organisational function, Adam's gesture also serves design-action management: refocusing topic, he aims to make his co-participants change design action.

### 4.4 Focalisation

Various examples of the focusing use of gestures have been presented above, in combination with designation and with organisation.

### 4.5 Modulating gestures

Extract 6 shows how Anna presents a book in which the author (A1, 34: "she") mentions the already "existing" crematorium. Anna had wanted to order the book, but encountered many administrative problems, so bought it herself. Telling this, she both highlights content elements (partly through beats) and expresses their emotional loading, using different types of gestures (deictics, iconics, metaphorics) (typographically coded in the extract). Pointing to the book (A1, 34), Anna uses the deictic gesture to highlight that "[they]'re mentioned in the book." Through frequently made long continuous hands opening, advancing metaphoric gestures (A1, 36, 37, 40, 41, 42, 43, 45-46), and a sequence of detached short beats (A1, 40-41), she underlines discourse components. Enacting how one holds a car steering wheel (A1, 46) and suddenly changes direction (A1, 47), she uses these iconic gestures to highlight her idea that people, normally, do not want to visit cemeteries during their holidays.

Extract 6, A1, Modulating discourse components using various forms of gesture
Caption *lchoag = long continuous hands opening, advancing gestures*

34 Anna yes she's been to have a look at our existing we're mentioned in the
[g_Anna.……………...
*points to the book*
35 book quite a bit with the existing chapel she was quite impressed
…....]
36 because we've also got quite a lot of photographs and other things and
[g_Anna.………………………….]
*long continuous hands opening, advancing gestures*
37 plans like the forward plan of extending the chapel originally the original
[g_Anna.………………………….……..]
*long continuous hands opening, advancing gestures*
38 idea so that was quite forward thinking in nineteen eighty or seventy
39 whenever seventy eight when they decided on that + so I've sent off for
40 that but you try and get them to raise a SAP order for it and you get how
[g_Anna.…………………………….…...…..] [g_Anna.……
*lchoag* *sequence of*
41 often will we need the SPIRE BOOKS COMPANY I said well we'll never use
………………………………………………] [g_Anna.………….]
*detached short beats* *lchoag*
42 them again probably why do you need to do that oh for god's sake
[g_Anna.……….]
*lchoag*
43 so I just went out and bought it and the one above as well a history
[g_Anna.………………..…..]
*lchoag*
44 of cremations that's what we have at home on the bookshelf
45 cheerful reading like you when you go on holiday with your part- your
[g_An

*lchoag*
46 husband and wife + you see a sign for crematorium straight away
na……………...] [g_Anna…..]
*enacting how one*
47 ignoring them whether you're abroad or in this country the
[g_Anna…………………………………]
*holds a car steering wheel*
*enacting how one suddenly changes direction*
48 first thing you do straight away oh not another cemetery they go oh dear

### 4.6 Disambiguation gestures
While the co-participants are discussing cars and off-loading the coffins from the cars, Anna refers to "the doors." Given the car-centred character of the interaction sequence and the absence of any clue signalling that a new topic is being introduced, the doors of *the cars* are probably the reference expected by the other co-participants. Yet, Anna's gesturing over the doors of the *waiting area* representation on the plan probably disambiguates their actual reference.

## 5. Discussion

### 5.1 Attributing fixed functions to particular gesture forms
In this text, designation was presented as mostly performed through deictics, and specifying mostly through iconic and metaphoric illustrative gestures and emblems, but we saw that, for example, deictics can have other uses and specification can be expressed differently. Indeed, there is no bijection between gesture forms and functions.

Besides their identificatory, qualificatory, or comparative orientation, designating gestures also have a focusing function. Other instances of this multifaceted nature of gesture were presented, for example:

- Designating representational gestures having an interactional function.
- Deictic designating gestures also functioning metaphorically.
- Various types of gestures used to modulate one's discourse.

### 5.2 Designating or designing
Our presentation of designating gestures may have seemed to presuppose that the entities designated existed already before the meeting, designed by the architect who "simply" presents them to his clients during the meeting. The architect, however, enunciates, gestures, and draws entities that he may be designing while he is conveying them to his co-participants. It is difficult, if not impossible, especially for an external observer, to distinguish between entities that "existed" already and entities that are *being designed* while they are being designated, that is, *in their designation*. Designing is taking place continuously: the meeting is not an event where a finished project is being reported.

Regarding the use of the artefact by different users (crematorium personnel, funeral directors, family of the deceased), especially in space and in motion (people sit here, pass by there), architectural plans are not explicit, and thus provide no fixed instructions. It is well possible, for example, that the itineraries that Adam indicates as possible ways to enter the crematorium area (in Extract 2) are designed by him in response to and in interaction—thus collaboration—with Anna when she formulates the funeral directors' worries concerning the access route.

This observation concerning gesture is to be situated in a wider context. Resuming what we wrote in *The cognitive artifacts of designing* (Visser 2006a, p. 199), an essential part of collaborative design occurs in and *through* interaction. Indeed, "the different forms that interaction may take in collaborative design—especially, linguistic, graphical, gestural, and postural—are… not the simple *expression* and *transmission* (communication) of ideas previously developed in an internal medium (such as Fodor's 'language of thought'). They are more and of a different nature than the trace of a so-called 'genuine' design activity, which would be individual and occur internally, and which verbal and other forms of expression would allow sharing with colleagues."

Several colleagues adopt comparable positions. According to Luck (this volume), talk IS action: it brings about "real" actions in the world. McDonnell (this volume) focuses on conversation as a medium through which a design project advances. Oak (this volume) analyses how design-meeting participants create their roles through their talk.

Concerning gesture specifically, recent research shows that it not only conveys information to one's addressees, but also affects the gesturer: it may help a person think (Lozano and Tversky 2006)—and learn (Goldin-Meadow and Wagner 2005).

### 5.3 The generic character of the gesture functions identified in A1

In order to appreciate the generic nature of our results with respect to the possibly idiosyncratic characteristics of the analysed meeting, we also viewed a meeting regarding another type of design in another stage (E1, an early engineering-design meeting). As in A1, both representational and organisational gestures were used in E1. Yet, particularly important in E1—more than in A1—seemed the gestural enactment of motion. Interaction around the possible motions of E1's artefact (a particular type of pen) involved sequences of gesturally performed actions, often many small ones. Even if the use of both a building and a pen imply action, it may be harder to formulate verbally and easier to express gesturally the many, small movements involved in manipulating a pen, compared to the different ways of passing through a crematorium.

Their correspondence with the results of previous studies on design meetings (cf. section 2.2) also substantiates the generic character of our results regarding at least two points, that is, activities performed using gesture and their objects. Gesture plays a role in both representational and organisational activities. They are used to specify both characteristics of design entities and aspects of user experience.

### 5.4 Irreplaceability of gesture relative to other semiotic modalities

Previous work on gesture in collaborative design confirmed the conclusion of classical gesture research (Kendon 2004, McNeill 2000) that gesture has a unique role and often cannot be replaced by other semiotic modalities.

Anticipating the next stage in our longitudinal project, we start a discussion of the degree to which gesture is irreplaceable relative to other semiotic systems.

Certain gestures in A1 seem to add nothing to the verbal expression: for example, Adam's hands moving apart when he asks how wide the passage would need to be for two coffins (Extract 3). Sometimes, a gesture has a function that might—also or better—be performed by a drawing. The possibility for people, when seated in the chapel, to see the coffin on the catafalque is a question of measuring on a drawing—even if this can be approximated by simulating the orientation of gaze using a finger. This does not imply, however, that questions concerning orientation of gaze can systematically be dealt with graphically (see Extract 4).

There are, conversely, many situations in which the contribution of gesture is not redundant relative to verbal or graphical expression. In the example presented in Section 4.6, *Disambiguation gestures*, it is only thanks to Anna's gesturing that her co-designers probably understand "the doors" as being the *waiting area* doors. People who would only *hear* her talk most probably would interpret them as *the car* doors.

Our preliminary conclusion is that, among the families of gestures identified in our study, gestures are neither systematically irreplaceable, nor "just" an optional accessory to speech or drawing. Furthermore, even if certain gestures might seem optional relative to other forms of expression, their discretionary character would require empirical confirmation.

The usefulness of gesture for the gesturer introduces another viewpoint on gesture's irreplaceability. A gesture that may be of little or no particular value for an addressee may play a unique role for the gesturer.

**5.5 Gesture in computer-supported collaborative software (CSCW) systems**

How should gestures that have been identified as irreplaceable in non-mediated collaboration, be handled in computer-supported collaborative software (CSCW) systems? CSCW-related research on gesture and gesture-substitution devices mostly focuses on technological aspects, without (reference to) empirical research on human gesturing in face-to-face interaction. When gesturing is supported, it is mostly deictic manipulation through pen-based devices, generally not other types of gesture, such as specifying.

In face-to-face interaction, designating gestures are often particularly useful associates of verbal expressions. If a verbal expression alone ("this…," "over there") is incomprehensible without a deictic gesture, can one use another mode (today, typically pen-based devices or telepointers), even if the substitute is not as easy and efficient as a deictic gesture?

Both analysis of face-to-face collaborative work, and specification and evaluation of technological approaches still require much endeavour in order to develop support systems that will be useful and useable in collaborative design meetings.

**5.6 Contribution to gesture studies**

Gesture has hardly been studied in collaborative professional activities (but cf. section 2.2). Most research has been conducted in interactional conditions that were not goal-oriented—generally, in conversational situations.

The present study has examined the function of gesture in a goal-oriented, professional work context, that is, a collaborative design meeting. It has shown how the "classical" gestures (such as deictics, iconics, metaphorics, and beats) are used with specific functions due to the task in which they were used, that is design.

We have identified representational and organisational gestures, corresponding to design activities' main functions. Organisational gestures occur in each activity, but representational gestures are, in our view, typical for design (Visser 2006a). Among the representational gestures, we set apart designation and specification. Specification gestures may seem *the* distinctive design gestures. However, if designation occurs in all interaction, it does not with the specific design function identified here. Concerning specification, all interaction involves specifying meaning for others in order to establish common ground, but, in design, specification aims not only understanding, but also agreement (a distinction Visser 2006b introduced in her analysis of "common ground").

### 5.7 Contribution to cognitive design research

The role of gesture as an irreplaceable semiotic system for expression is not restricted to design: people not only always gesture—be they alone or in interaction, blind or sighted, visible for their addressees or not—, they would not be able to express themselves satisfactorily without. For speaking and hearing people, speech and gesture operate as an inseparable unit, reflecting different semiotic aspects of the cognitive structure that underlies both (McNeill 2000). What is under discussion here is the specific contribution of gesture to collaborative design.

This study contributes to our knowledge on design thinking as regards both the activities in design and the bodily realisation of these activities. As to aspects related to space (especially 3D), motion, and (sequences of) action, it points out (1) the particular nature of the way in which they are being designed and (2) the semiotic system of gesture as especially prone to the implementation of these design activities. Our results indeed point to the essential role of gesture in both the design of, and the communication around 3D and dynamic design aspects (such as motion and action), which are difficult, if not impossible, to elaborate using verbal or graphical expression (or mock-ups, for the dynamic design aspects).

Until now, cognitive design studies have been silent concerning designational activities. Therefore, was it not the present analysis, we would most probably have remained unaware of the magnitude of these activities in collaborative design. We knew gesture's appropriateness to render signalling through deictics. What our results add to this knowledge are (1) the importance of designational activities with a design function in collaborative design, and (2) the implementation of this function through gesture.

Even if it is not specific to design, we also wish to underline the importance of gesture—a "silent" form of expression—for organisational matters. This is indeed a noteworthy result, as it confirms previous work on gesture in design meetings. The present study, however, has shown this organisational role of gesture to have a wider coverage than attributed in other research, including also the management of design actions.

## Acknowledgements

The author gratefully acknowledges Françoise Détienne's remarks concerning the first version, and the comments and suggestions by two anonymous colleagues concerning the second version of this text.